\documentclass[reprint,superscriptaddress,amsmath,amssymb,aps]{revtex4-2}

\usepackage[pdftex]{graphicx}
\usepackage[pdftex]{xcolor}
\usepackage[pdftex,breaklinks,colorlinks=true,
            linkcolor=blue,citecolor=blue,urlcolor=magenta]{hyperref}

\begin{document}

\title{
Electric-field-induced magnetic toroidal moment and nonlinear magnetoelectric effect in antiferromagnetic olivines
}

\author{Yasuyuki~Kato}
\email{katoyasu@u-fukui.ac.jp}
\affiliation{Department of Applied Physics, University of Fukui, Fukui 910-8507, Japan}

\author{Takeshi~Hayashida}
\affiliation{Department of Applied Physics, The University of Tokyo, Tokyo 113-8656, Japan}
\affiliation{FELIX Laboratory, Radboud University, Toernooiveld 7, 6525 ED Nijmegen, the Netherlands}

\author{Koei~Matsumoto}
\affiliation{Department of Applied Physics, The University of Tokyo, Tokyo 113-8656, Japan}

\author{Tsuyoshi~Kimura}
\affiliation{Department of Applied Physics, The University of Tokyo, Tokyo 113-8656, Japan}

\author{Yukitoshi~Motome}
\affiliation{Department of Applied Physics, The University of Tokyo, Tokyo 113-8656, Japan}

\date{\today}

\begin{abstract}
Beyond conventional electric and magnetic monopoles, electric and magnetic toroidal monopoles, 
which are rank-0 multipoles distinguished by opposite parities under spatial inversion and time reversal, 
can exist in nature. 
The recent observation of electric-field-induced directional dichroism in antiferromagnetic olivine Co$_2$SiO$_4$ 
has provided the first concrete example of a magnetic toroidal monopole; 
however, its microscopic origin remains elusive. 
Here, we propose a minimal spin model that incorporates magnetoelectric coupling via the $d$-$p$ hybridization mechanism 
and analyze it within the mean-field approximation. 
The model qualitatively reproduces the experimentally observed temperature dependence of the dielectric constant 
and its pronounced sensitivity to the direction of the applied electric field. 
Furthermore, it elucidates the temperature evolution of the magnetic toroidal monopole 
and the strong electric-field-direction dependence of the magnetic toroidal moment. 
Our calculations also predict a second-order nonlinear magnetoelectric response,
consistent with the symmetry classification of Co$_2$SiO$_4$ as an altermagnet.
Additionally, we demonstrate that the same framework is applicable to other antiferromagnetic olivines 
with analogous magnetic order, indicating the robustness and generality 
of the toroidal-type magnetoelectric response in this material family.
\end{abstract}

\maketitle

\section{Introduction} \label{sec:introduction}
From the viewpoint of two fundamental symmetries in physics, spatial inversion and time reversal, 
the well-known electric and magnetic monopoles share the same parity combinations with respect to these symmetry operations: 
the electric monopole is even under both, while the magnetic monopole is odd under both.
In contrast, monopoles with opposite parity combinations can also exist. 
A monopole that is even under spatial inversion and odd under time reversal is referred to as a \textit{magnetic toroidal monopole}, 
while one that is odd under spatial inversion and even under time reversal is an \textit{electric toroidal monopole}~\cite{Kusunose2022,Hayami2025}.
Within the multipole framework,
monopoles represent the lowest order in the hierarchy of multipoles as rank-0 scalar quantities and serve as prototypes for higher-rank multipoles.
Given the fundamental role played by electric and magnetic monopoles in physics, 
it is crucially important to establish a systematic understanding of magnetic and electric toroidal monopoles. 

In this study, we focus on the magnetic toroidal monopole.
From a symmetry perspective, it can couple to both an electric field $\mathbf{E}$ 
(even under spatial inversion and odd under time reversal, corresponding to a rank-1 electric multipole field) and 
a magnetic toroidal moment $\mathbf{t}$ 
(odd under both spatial inversion and time reversal, corresponding to a rank-1 magnetic toroidal multipole,
i.e., the magnetic toroidal dipole). 
Consequently, the free energy of the system may include a coupling term 
such as $T_0 \mathbf{E} \cdot \mathbf{t}$, where $T_0$ denotes the magnetic toroidal monopole.
Therefore, 
in the presence of nonzero $T_0$, the system can exhibit
a peculiar response: a magnetic toroidal moment is induced by an applied electric field~\cite{Hayami2023}. 

Recently, electric-field-induced directional dichroism has been observed in the antiferromagnetic olivine Co$_2$SiO$_4$, 
suggesting the presence of a magnetic toroidal monopole~\cite{Hayashida2025}. 
The magnetic point group of the antiferromagnetic state in this compound is $mmm$, which allows a finite $T_0$ from a symmetry viewpoint~\cite{Hayami2023}. 
Notably, an antiferromagnet with $mmm$ symmetry is classified as an altermagnet, characterized by the breaking both 
$\mathcal{T}$ and $\mathcal{PT}$ symmetries, 
where $\mathcal{T}$ and $\mathcal{P}$ represent time reversal and space inversion,
respectively~\cite{Yuan2021,Smejkal2022a,Smejkal2022b,Cheong2025}. 
The observed directional dichroism implies that 
an electric field induces a magnetic toroidal moment~\cite{Xu2024}, 
and thus, the peculiar response can be qualitatively understood within the phenomenological framework based on the coupling term described above. 
However, its microscopic origin, particularly the roles of spin degrees of freedom and orbital hybridization,
remains unclear. 
Despite their importance for confirming the existence of the magnetic toroidal monopole in this material and advancing our understanding of the underlying mechanism, 
a comprehensive microscopic description has yet to be established.

\begin{figure*}[!tbh]
\centering
\includegraphics[trim = 0 0 0 0, width=\textwidth,clip]{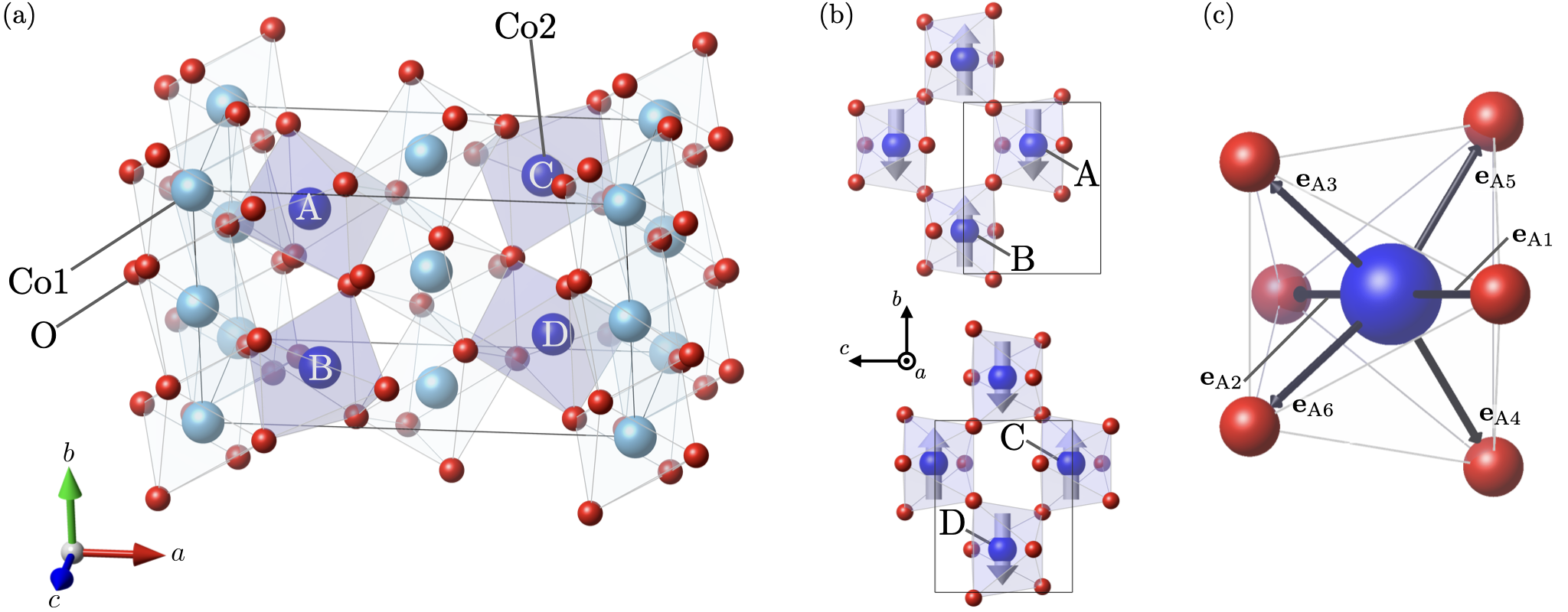}
\caption{\label{fig:fig01}
Crystal structure of Co$_2$SiO$_4$~\cite{Sazonov2008}.  
For clarity, Si atoms are omitted, showing only Co and O. 
(a) The unit cell (thin black lines) contains two crystallographically distinct Co$^{2+}$ sites, 
denoted as Co1 (light blue) and Co2 (dark blue). 
Each Co$^{2+}$ ion is octahedrally coordinated by six O$^{2-}$ ligands (red). 
In this study, we focus on the Co2 sites and assign sublattice indices A, B, C, and D within the unit cell. 
(b) The Co2 sublattices A and B, and C and D, respectively, form square-lattice-like networks in the $bc$ plane 
through corner sharing of octahedra. 
Arrows indicate the spin directions (parallel or antiparallel to the $b$ axis) in the ordered phase. 
(c) Octahedron formed by a Co2 ion of sublattice A and its surrounding ligands. 
$\mathbf{e}_{\mathrm{A}1}$--$\mathbf{e}_{\mathrm{A}6}$ represent the relative position vectors from the Co2 site to the ligands.
Crystal structures are visualized using VESTA~\cite{Momma2011}.
}
\end{figure*}

In this paper, we propose a minimal model to describe
the electric-field-induced magnetic toroidal moment in Co$_2$SiO$_4$ and clarify its microscopic origin. 
Based on the local environment of Co$^{2+}$ ions and the $d$-$p$ hybridization mechanism~\cite{Arima2007}, 
we introduce a spin model with magnetoelectric coupling and analyze it within the mean-field approximation. 
We show that this minimal model qualitatively reproduces the experimentally observed temperature dependence of the dielectric constant for different directions of the electric field. 
Furthermore, we theoretically reveal the temperature dependence of the magnetic-toroidal-monopole density, 
the electric-field-direction dependence of the magnetic toroidal moment,
and an antisymmetric magnetoelectric response associated with the magnetic toroidal monopole, which we refer to as
the nonlinear electric-field-induced toroidal-type magnetoelectric effect. 
Our findings provide not only experimentally measurable evidence for the existence of magnetic toroidal monopoles 
but also a microscopic understanding of toroidal-type magnetoelectric responses in antiferromagnetic olivines. 

The remainder of this paper is organized as follows. 
In Sec.~\ref{sec:model_and_method}, we define the spin model, 
introduce the electric-polarization operator based on the $d$-$p$ hybridization mechanism, and describe the 
method used for the following analyses. 
The corresponding experimental procedure is briefly described in the same section. 
Section~\ref{sec:results_and_discussion} presents the main results: 
a comparison between theoretical and experimental dielectric anomalies~(Sec.~\ref{sec:resultA}), 
the temperature dependence of the magnetic-toroidal-monopole density~(Sec.~\ref{sec:resultB}), 
the electric-field-direction dependence of the magnetic toroidal moment~(Sec.~\ref{sec:resultC}), 
and the second-order magnetoelectric response, i.e., 
the electric-field-induced toroidal-type magnetoelectric effect~(Sec.~\ref{sec:resultD}). 
In addition, we extend the same framework to other antiferromagnetic olivines with similar magnetic order and compare the results~(Sec.~\ref{sec:resultE}). 
Finally, Sec.~\ref{sec:summary}
summarizes the key findings and discusses possible future perspectives.

\section{Model and methods}\label{sec:model_and_method}
\subsection{Theoretical approach}\label{sec:theory}

Figure~\ref{fig:fig01}(a) shows the crystal structure of Co$_2$SiO$_4$. 
Each unit cell contains eight magnetic Co$^{2+}$ ions with $S=3/2$, 
which are crystallographically classified into two inequivalent sites, 
referred to as Co1 and Co2, each hosting four ions.
Each Co$^{2+}$ ion is octahedrally coordinated by six O$^{2-}$ ligands,
and the octahedra share corners or edges to form a three-dimensional network. 
Specifically,
the Co1 octahedra form one-dimensional edge-sharing chains along the $b$ axis, 
while the Co2 octahedra constitute corner-sharing square-lattice-like networks within the $bc$ plane, 
as shown in Fig.~\ref{fig:fig01}(b).
This compound undergoes a magnetic phase transition at approximately 50~K~\cite{Lottermoser1988,Ballet1989,Sazonov2009}.
In the magnetically ordered phase, the Co2 spins form a collinear antiferromagnetic structure along the $b$ axis as shown in Fig.~\ref{fig:fig01}(b), 
while the magnetic moments on the Co1 sites exhibit a noncoplanar configuration slightly tilted from the $b$ axis. 
Because including the Co1 spins makes the model unnecessarily complex, 
we focus solely on the Co2 sites in the following analysis.

The model considered here consists of the nearest-neighbor antiferromagnetic Heisenberg interaction $J$
on the square-lattice-like network of Co2 ions,  
the single-ion anisotropy $A$ favoring an easy axis along the $b$ direction,
and the Zeeman and magnetoelectric coupling terms representing the interactions of spins
with external magnetic and electric fields, respectively.  
The Hamiltonian is given by
\begin{align}
\mathcal{H} = J \sum_{\langle \mathbf{r}, \mathbf{r}' \rangle} \mathbf{S}_{\mathbf{r}} \cdot \mathbf{S}_{\mathbf{r}'} 
- A \sum_{\mathbf{r}} (S_{\mathbf{r}}^{b})^{2} 
- \mathbf{E} \cdot \sum_{\mathbf{r}} \mathbf{p}_{\mathbf{r}} 
- \mathbf{h} \cdot \sum_{\mathbf{r}} \mathbf{S}_{\mathbf{r}},
\label{eq:H}
\end{align}
where $\mathbf{S}_{\mathbf{r}}$ is the spin operator at site $\mathbf{r}$ with $S=3/2$ 
($S_{\mathbf{r}}^{b}$ represents its component along the $b$ axis), 
$\sum_{\mathbf{r}}$ and $\sum_{\langle \mathbf{r}, \mathbf{r}' \rangle}$ denote summations over all sites and nearest-neighbor pairs, respectively, 
$\mathbf{E}$ and $\mathbf{h}$ are respectively the external electric and magnetic fields, 
and $\mathbf{p}_{\mathbf{r}}$ represents the local spin-dependent electric polarization operator defined below.

To incorporate the magnetoelectric effect in this material, 
we consider the $d$-$p$ hybridization mechanism arising from the mixing between Co$^{2+}$ $d$ orbitals and O$^{2-}$ $p$ orbitals~\cite{Arima2007}.
This mechanism becomes relevant only when the transition-metal site lacks an inversion center. 
This condition is indeed the case for Co2, as illustrated in Fig.~\ref{fig:fig01}(c). 
For a Co2 ion belonging to sublattice $\alpha$ (A, B, C, or D), 
the relative position vectors of the six ligands are denoted by $\mathbf{e}_{\alpha\mu}$ ($\mu=1,2,\dots,6$) 
and the corresponding unit vectors by $\hat{\mathbf{e}}_{\alpha\mu}$. 
The local electric-polarization operator in the $d$-$p$ hybridization mechanism is then defined as
\begin{align}
\mathbf{p}_{\mathbf{r}} = \sum_{\mu} \lambda_{\alpha(\mathbf{r})\mu}
\left( \mathbf{S}_{\mathbf{r}} \cdot \hat{\mathbf{e}}_{\alpha(\mathbf{r})\mu} \right)^{2}
\hat{\mathbf{e}}_{\alpha(\mathbf{r})\mu},
\end{align}
where $\alpha(\mathbf{r})$ specifies the sublattice of site $\mathbf{r}$. 
The coefficient $\lambda_{\alpha\mu}$ depends on the crystal field and the strength of $d$-$p$ hybridization. 
Although it is generally sublattice dependent, we set $\lambda_{\alpha\mu}=1$ for simplicity.

We investigate the model in Eq.~\eqref{eq:H} using the mean-field approximation.
Within this framework,
the spin-exchange term is decoupled as
\begin{align}
\mathbf{S}_{\mathbf{r}} \cdot \mathbf{S}_{\mathbf{r}'}
\simeq
\mathbf{S}_{\mathbf{r}} \cdot \mathbf{m}_{\alpha(\mathbf{r}')}
+ \mathbf{m}_{\alpha(\mathbf{r})} \cdot \mathbf{S}_{\mathbf{r}'}
- \mathbf{m}_{\alpha(\mathbf{r})} \cdot \mathbf{m}_{\alpha(\mathbf{r}')},
\end{align}
with the mean field 
$\mathbf{m}_{\alpha(\mathbf{r})}=\langle\mathbf{S}_{\mathbf{r}}\rangle$, 
which is computed as the thermal average within the mean-field Hamiltonian with this decoupling and is determined self-consistently. 
In the present model, although the sublattice pairs (A,B) and (C,D) are decoupled across different $bc$ planes, 
a mean-field solution consistent with the experimental magnetic order can be obtained 
by initializing the configuration as shown in Fig.~\ref{fig:fig01}(b).

In the following calculations, we set $J=1$ as the energy unit and take the single-ion anisotropy $A=0.1$. 
We confirm that calculations with $A=0.01$ and $A=1$ give qualitatively similar results. 
This indicates that the physical quantities discussed in this study are rather insensitive to the value of $A$, 
indicating the robustness of our results.

\subsection{Experimental approach}

A single crystal of Co$_2$SiO$_4$ was grown by the floating zone method~\cite{Tang2011}. 
The obtained crystal was oriented using Laue X-ray diffraction, 
and three plate-shaped samples with the widest faces perpendicular to the $a$, $b$, and $c$ axes were prepared. 
The dimensions of the samples were approximately $3 \mathrm{mm} \times 3 \mathrm{mm} \times 0.2 \mathrm{mm}$. 
Magnetization measurements confirmed that the samples exhibit antiferromagnetic order at $50$ K. 
For dielectric measurements, silver paste was painted onto the widest faces of the samples to form a pair of electrodes. 
The dielectric constant along the principal crystallographic axes was measured using an LCR meter (E4980A, Agilent) 
while the temperature was controlled with a commercial physical property measurement system (PPMS, Quantum Design).

\section{Results and Discussion}\label{sec:results_and_discussion}
\subsection{Dielectric anomalies}\label{sec:resultA}

\begin{figure*}
\centering
\includegraphics[trim = 0 0 0 0, width=\textwidth,clip]{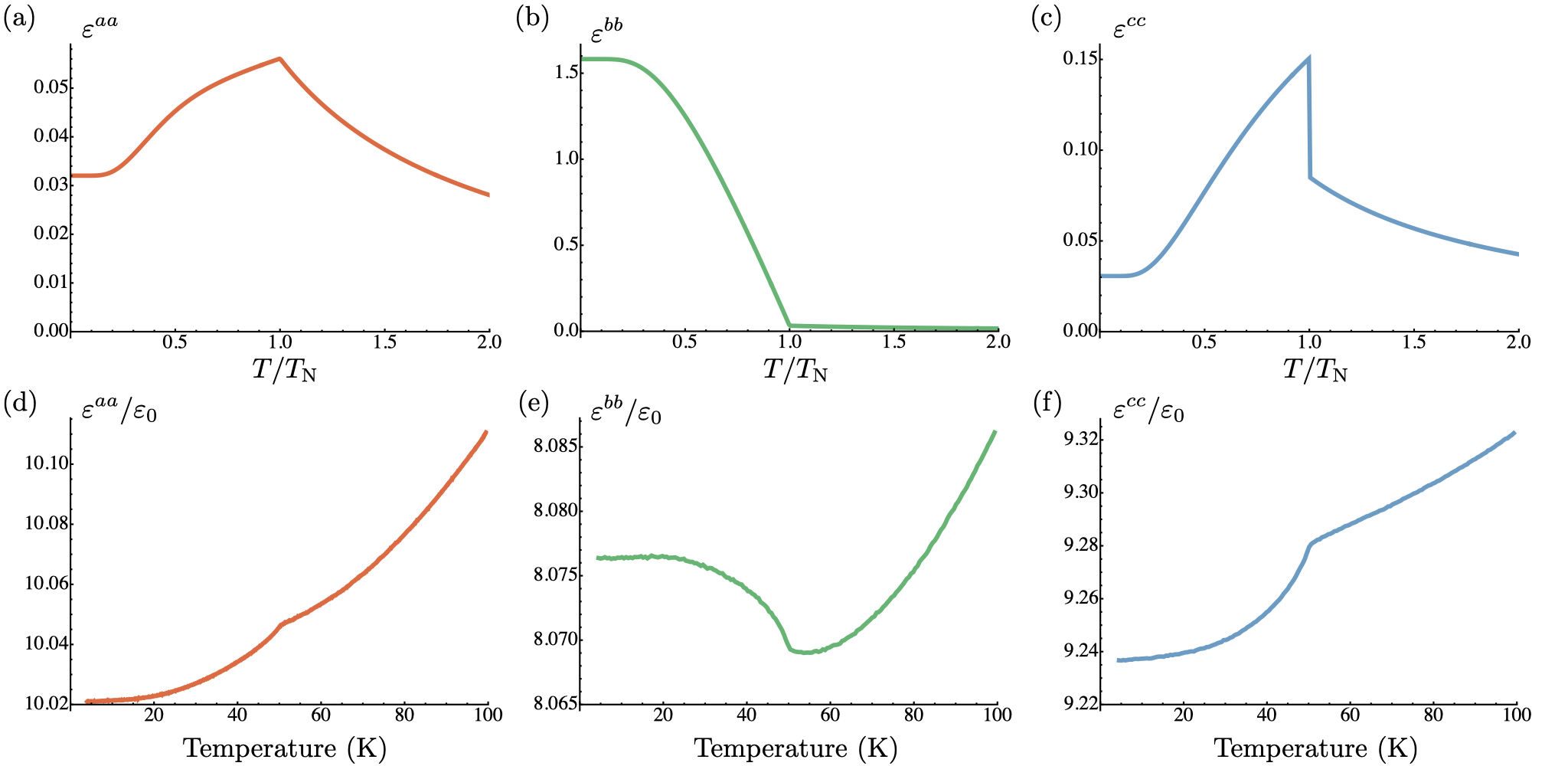}
\caption{\label{fig:fig02}
Temperature dependence of the dielectric constant from theory and experiment. 
Panels (a)--(c) show the theoretical results of (a) $\varepsilon^{aa}$, (b) $\varepsilon^{bb}$, and (c) $\varepsilon^{cc}$. 
Panels (d)--(f) show the corresponding experimental data of (d) $\varepsilon^{aa}$, (e) $\varepsilon^{bb}$, and (f) $\varepsilon^{cc}$. 
}
\end{figure*}

Figure~\ref{fig:fig02} compares the temperature ($T$) dependence of the dielectric constant along the principal crystallographic axes obtained from both 
theoretical calculations for the model in Eq.~\eqref{eq:H} and experimental measurements for Co$_2$SiO$_4$. 
In the theoretical calculations, the electric polarization per Co2 site is defined as
\begin{align}
\mathbf{p} = \frac{1}{4}\left(
\mathbf{p}_{\mathrm{A}} + \mathbf{p}_{\mathrm{B}} + \mathbf{p}_{\mathrm{C}} + \mathbf{p}_{\mathrm{D}}
\right),
\end{align}
where $\mathbf{p}_{\alpha}$ represents the thermal expectation value of the local spin-dependent polarization operator, $\langle\mathbf{p}_{\mathbf{r}}\rangle$, 
on the Co2 sites belonging to sublattice $\alpha$.
We probe the linear response by applying a weak electric field 
along each crystallographic axis $\mu=a,b,c$, i.e.,~$\mathbf{E}=E^{\mu}\hat{\mathbf{e}}_{\mu}$. 
The dielectric constant is defined as the directional derivative
\begin{align}
\varepsilon^{\mu\mu}=\left.\frac{\partial p^{\mu}}{\partial E^{\mu}}\right|_{\mathbf{E}=0},
\end{align}
which is numerically evaluated using
\begin{align}
\varepsilon^{\mu\mu}\approx
\frac{ 
\left. p^{\mu} \right|_{\mathbf{E} = E^{\mu}\hat{\mathbf{e}}_{\mu}}
-\left. p^{\mu} \right|_{\mathbf{E} = 0}
}{E^{\mu}},
\end{align}
with a sufficiently small field strength $E^\mu=0.005$. 
Due to the symmetry of the system, 
the polarization vanishes at zero field, i.e., 
$\mathbf{p}=\mathbf{0}$ at $\mathbf{E}=\mathbf{0}$,
as discussed in Sec.~\ref{sec:resultB}. 
Therefore, the above expression reduces to 
$\varepsilon^{\mu\mu}\approx 
\left. p^{\mu} \right|_{\mathbf{E} = E^{\mu}\hat{\mathbf{e}}_{\mu}}
/E^{\mu}$.

The theoretical results shown in Figs.~\ref{fig:fig02}(a)--\ref{fig:fig02}(c) exhibit a pronounced dependence on the field direction, 
while all results consistently show an anomaly at the N\'eel temperature $T_{\mathrm{N}} \approx 5.08$. 
Specifically, convex anomalies are observed for the $a$ and $c$ axes, 
whereas a concave anomaly appears for the $b$ axis.
Moreover, in the ordered phase ($T < T_{\mathrm{N}}$), $ \varepsilon^{aa}$ and $ \varepsilon^{cc}$ decrease monotonically with decreasing temperature, while $ \varepsilon^{bb}$ increases. 
The experimental data presented in Figs.~\ref{fig:fig02}(d)--\ref{fig:fig02}(f) show trends that are qualitatively reproduced by the theoretical results [Figs.~\ref{fig:fig02}(a)--\ref{fig:fig02}(c)], 
although they likely include monotonically increasing background contributions with temperature. 
Since no established method is available to subtract these contributions, 
the raw data are presented as measured.
In contrast, the theoretical calculation shows a discontinuous jump in $ \varepsilon^{cc}$ at $ T_{\mathrm{N}}$, 
which is not observed experimentally. 
This discrepancy is likely an artifact of the mean-field approximation and the simplifications inherent to the current model.

Taken together, these results indicate that the proposed minimal model qualitatively reproduces the experimentally observed temperature and field-direction dependences of the dielectric constant. 
The consistency between the theoretical and experimental trends indicates that the present model provides a reasonable microscopic description of the magnetoelectric behavior in Co$_2$SiO$_4$.

\subsection{Magnetic toroidal monopole at zero fields}\label{sec:resultB}
\begin{figure}[!htbp]
\centering
\includegraphics[trim = 0 0 0 0, width=\columnwidth,clip]{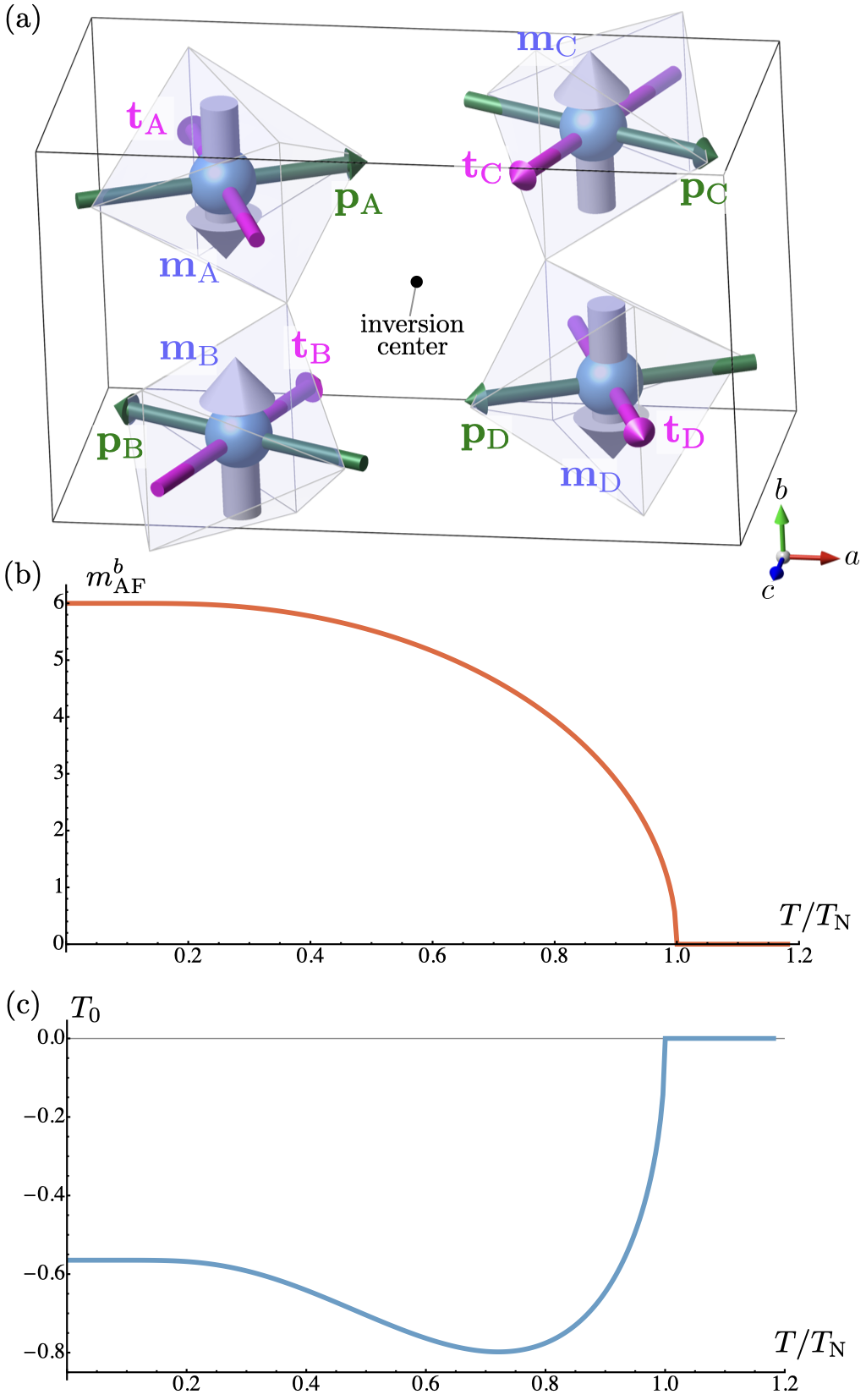}
\caption{\label{fig:fig03}
Antiferromagnetic order and magnetic toroidal monopole. 
(a) Real-space distributions of the local magnetization $\mathbf{m}_\alpha$, 
the local electric polarization $\mathbf{p}_\alpha$, 
and the local magnetic toroidal moment $\mathbf{t}_\alpha$ 
in the low-temperature antiferromagnetically ordered phase. 
(b) Temperature dependence of the antiferromagnetic order parameter $m_{\mathrm{AF}}^b$. 
(c) Temperature dependence of the magnetic toroidal monopole $T_0$.}
\end{figure}

Figure~\ref{fig:fig03}(a) displays the spatial configurations of the local magnetization $\mathbf{m}_\alpha$ 
and the local electric polarization $\mathbf{p}_\alpha$, 
obtained from the mean-field calculation below $T=T_{\rm N}$ at $\mathbf{E}=0$ and $\mathbf{h}=0$.
In the mean-field solution, all Co2 sites carry magnetic moments of the same magnitude, 
with $\mathbf{m}_\mathrm{A} = \mathbf{m}_\mathrm{D}$ pointing along the negative $b$ direction 
and $\mathbf{m}_\mathrm{B} = \mathbf{m}_\mathrm{C}$ along the positive $b$ direction. 
Each Co2 ion possesses a mirror plane perpendicular to the $b$ axis (the $ac$ plane), 
which forces the $b$ component of the local electric polarization to vanish ($p_\alpha^b = 0$). 
The A and D sublattices are related by a 180$^\circ$ rotation about the $b$ axis, 
up to an appropriate lattice translation,
yielding $\mathbf{p}_\mathrm{A} = -\mathbf{p}_\mathrm{D}$. 
Similarly, the B and C sublattices satisfy $\mathbf{p}_\mathrm{B} = -\mathbf{p}_\mathrm{C}$. 
Furthermore, the A and B sublattices are transformed into each other by reflection across the $bc$ plane,
again up to an appropriate lattice translation. 
This operation reverses only the $a$ component of the polarization while keeping the $c$ component identical, 
namely $p_\mathrm{A}^a = -p_\mathrm{B}^a$ and $p_\mathrm{A}^c = p_\mathrm{B}^c$. 
Similar relations also hold between B and D sublattices: $p_\mathrm{C}^a = -p_\mathrm{D}^a$ and $p_\mathrm{C}^c = p_\mathrm{D}^c$.
As a result of these three symmetry operations, the $ac$-plane mirror, 
the 180$^\circ$ rotation about the $b$ axis, 
and the $bc$-plane mirror, the net electric polarization vanishes.

The antiferromagnetic order shown in Fig.~\ref{fig:fig03}(a) 
is characterized by 
\begin{align}
\mathbf{m}_{\mathrm{AF}} = 
 - \mathbf{m}_\mathrm{A} + \mathbf{m}_\mathrm{B}
 + \mathbf{m}_\mathrm{C} - \mathbf{m}_\mathrm{D},
\end{align}
whose $b$ component $m_{\mathrm{AF}}^b$ serves as the conventional order parameter. 
As shown in Fig.~\ref{fig:fig03}(b), $m_{\mathrm{AF}}^b$ develops below the N\'eel temperature $T_{\mathrm{N}}$
and exhibits a typical square-root-type $T$ dependence near $T_{\mathrm{N}}$.

The noncollinear configuration of $\mathbf{m}_\alpha$ and $\mathbf{p}_\alpha$ in the antiferromagnetically ordered phase activates a magnetic toroidal moment at each Co2 site. 
Following Refs.~\cite{Kimura2024,Hayashida2025}, 
the local magnetic toroidal moment is defined as
\begin{align}
\mathbf{t}_\alpha = \mathbf{p}_\alpha \times \mathbf{m}_\alpha,
\end{align}
which, by definition, is perpendicular to both $\mathbf{m}_\alpha$ and $\mathbf{p}_\alpha$. 
The same symmetry relations that apply to $\mathbf{p}_\alpha$ also hold for $\mathbf{t}_\alpha$:
$\mathbf{t}_\mathrm{A} = -\mathbf{t}_\mathrm{D}$, 
$\mathbf{t}_\mathrm{B} = -\mathbf{t}_\mathrm{C}$, 
$t_\mathrm{A}^a = -t_\mathrm{B}^a$, and $t_\mathrm{A}^c = t_\mathrm{B}^c$. 
The resultant $\mathbf{t}_\alpha$ are shown in Fig.~\ref{fig:fig03}(a). 
This peculiar spatial configuration of $\mathbf{t}_\alpha$ generates a magnetic toroidal monopole via the definition:
\begin{align}
T_0 = \frac{1}{3} \sum_{\alpha} \mathbf{r}_\alpha \cdot \mathbf{t}_\alpha,
\end{align}
where $\mathbf{r}_\alpha$ denotes the position vector of the Co2 ion in sublattice $\alpha$, 
measured from the inversion center of the unit cell shown by the black dot in Fig.~\ref{fig:fig03}(a).

Figure~\ref{fig:fig03}(c) presents the temperature dependence of $T_0$. 
Similar to $m_{\mathrm{AF}}^b$, 
it becomes finite below $T_{\mathrm{N}}$, while exhibiting nonmonotonic behavior at low temperatures. 
Although the magnitude of $T_0$ depends on the choice of the unit cell, 
its nonzero value is essential: 
$T_0 \neq 0$ implies that the coupling between the magnetic toroidal moment (dipole) and the electric field is symmetry-allowed. 
Notably, $T_0$ represents the lowest-rank (rank-0) scalar quantity among magnetic toroidal multipoles, 
forming the most fundamental element in this multipole series.
Our results clearly demonstrate the existence of the fundamental magnetic toroidal monopole associated with the antiferromagnetic order in this system.

\subsection{Electric-field-induced magnetic toroidal moment}\label{sec:resultC}

\begin{figure}[!htbp]
\centering
\includegraphics[trim = 0 0 0 0, width=\columnwidth,clip]{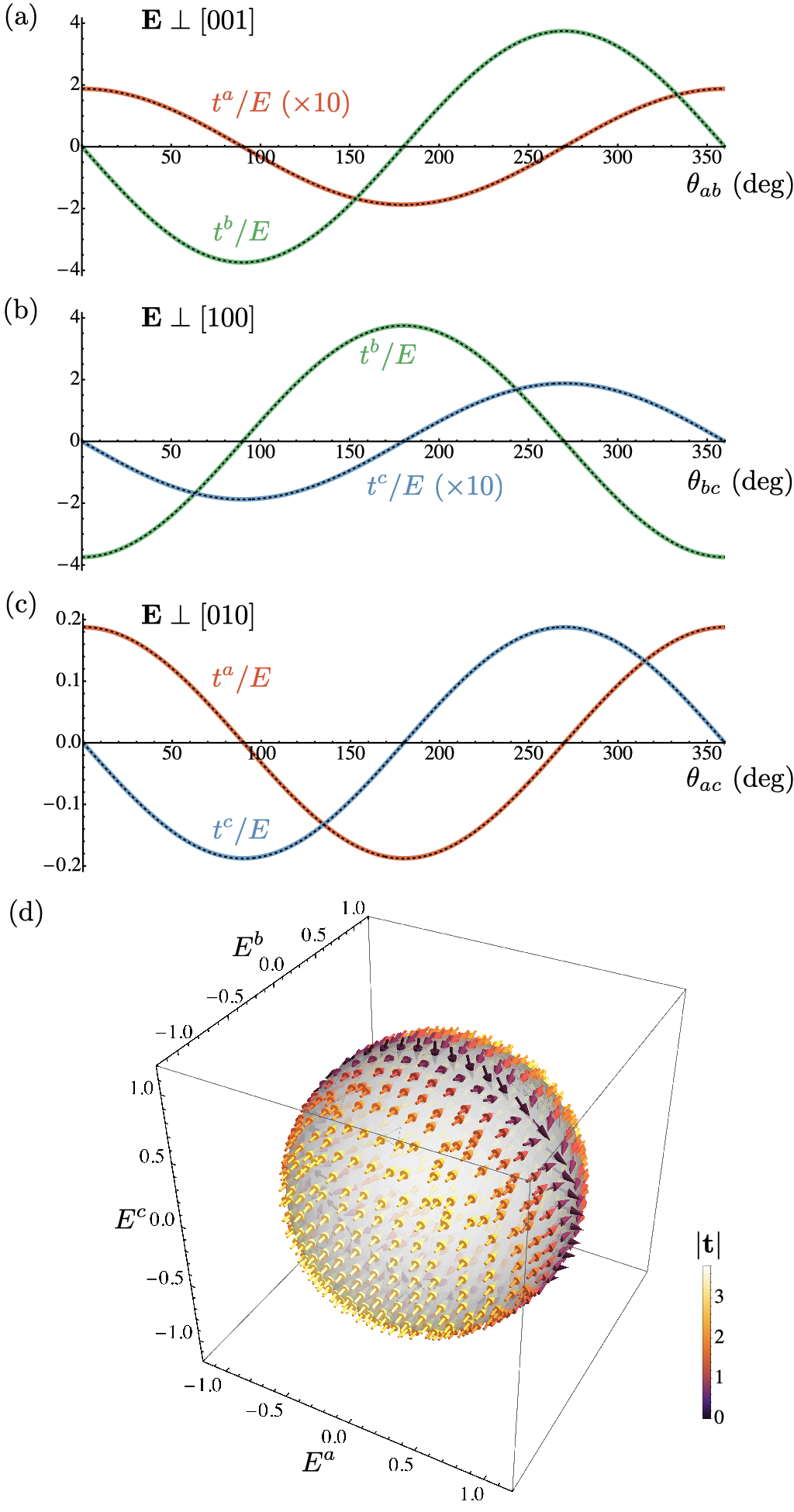}
\caption{\label{fig:fig04}
Electric-field-induced magnetic toroidal moment. 
(a)--(c) Field-angle dependence of the induced components:
(a) $t^a$ and $t^b$ for $\mathbf{E} \perp c$, 
(b) $t^b$ and $t^c$ for $\mathbf{E} \perp a$, and 
(c) $t^a$ and $t^c$ for $\mathbf{E} \perp b$. 
Solid lines are the numerical results obtained within the mean-field approximation,
and dashed lines are fits to either $\cos\theta$ or $\sin\theta$ functions.
(d) Distribution of the induced toroidal-moment response with respect to the electric-field direction, visualized on a sphere. 
The color of each arrow indicates the magnitude of the magnetic toroidal moment 
$|\mathbf{t}| = \sqrt{\sum_{\mu} (X^{tE}_{\mu\mu} E^{\mu})^2}$, 
where $X^{tE}_{\mu\mu}$ is the magnetoelectric coefficient estimated from the fitting, 
for an electric field with normalized amplitude $|\mathbf{E}|=1$. 
}
\end{figure}

When the magnetic toroidal monopole $T_0$ is finite, the electric field $\mathbf{E}$ can induce a net magnetic toroidal moment~\cite{Schmid2001,Schmid2008}. 
The observation of directional dichroism in Co$_2$SiO$_4$ strongly suggests the existence of this electric-field-induced magnetic toroidal moment~\cite{Hayashida2025}.
Within the linear response, this phenomenon is described by
\begin{align}
t^{\mu} = X^{tE}_{\mu\nu} E^{\nu},
\end{align}
where $t^{\mu}$ represents the $\mu$ component of the induced magnetic toroidal moment,
defined as
\begin{align}
\mathbf{t} = 
\mathbf{t}_{\mathrm{A}} + \mathbf{t}_{\mathrm{B}} +
\mathbf{t}_{\mathrm{C}} + \mathbf{t}_{\mathrm{D}},
\end{align}
and $X^{tE}_{\mu\nu}$ denotes
the linear magnetoelectric coefficient.

We analyze this effect within our model by applying a sufficiently weak electric field with magnitude $E = |\mathbf{E}| = 0.005$. 
Figures~\ref{fig:fig04}(a)--\ref{fig:fig04}(c) present the field-angle dependence of the induced $\mathbf{t}$ under $\mathbf{E}$. 
Here, the electric field is rotated within the $ab$, $bc$, and $ac$ planes as 
\begin{align}
\mathbf{E}_{ab} &= E (\cos\theta_{ab}, \sin\theta_{ab}, 0), \label{eq:Eab}\\
\mathbf{E}_{bc} &= E (0, \cos\theta_{bc}, \sin\theta_{bc}), \label{eq:Ebc}\\
\mathbf{E}_{ac} &= E (\cos\theta_{ac}, 0, \sin\theta_{ac}). \label{eq:Eac}
\end{align}
Each induced component $t^{\mu}$ is well fitted by either a $\cos\theta$ or $\sin\theta$ function (dashed lines), 
enabling extraction of the coefficients $X^{tE}_{\mu\nu}$. 
The results reveal that only the diagonal components of $X^{tE}_{\mu\nu}$ are nonzero:
\begin{align}
X^{tE}_{aa} = 0.188, \quad
X^{tE}_{bb} = -3.748, \quad
X^{tE}_{cc} = -0.188 .
\end{align}
Thus, the linear response is most pronounced along the antiferromagnetic moment direction $b$, accompanied by much weaker responses in the $a$ and $c$ directions, with $X^{tE}_{aa} = -X^{tE}_{cc}$. 
The relation between these coefficients in the weaker responses is likely not coincidental and can be understood as follows.  
At $\mathbf{E} = \mathbf{0}$, the magnetic moment $\mathbf{m}_\alpha$ has only the $b$ component, so that $\mathbf{t}_\alpha$ is given by
\begin{align}
t^a_\alpha = -p^c_\alpha m^b_\alpha, \quad
t^b_\alpha = 0, \quad
t^c_\alpha =  p^a_\alpha m^b_\alpha.
\end{align}
Within the linear response regime, 
the local electric polarization $\mathbf{p}_\alpha$ varies linearly with $\mathbf{E}$, 
and its change can be expressed using a ``local dielectric tensor'' $\varepsilon_\alpha$ as $\Delta \mathbf{p}_\alpha = \varepsilon_\alpha \mathbf{E}$.
Using this expression, the $E^\mu$-induced change in $t^\mu_\alpha$ is written as
\begin{align}
\Delta t^a_\alpha = - \varepsilon_{\alpha}^{ca} E^a m^b_\alpha,\quad 
\Delta t^b_\alpha = 0,\quad 
\Delta t^c_\alpha = \varepsilon_{\alpha}^{ac} E^c m^b_\alpha.
\end{align}
Since $\mathbf{p}_{\alpha}$ is even under time reversal,
Onsager reciprocity requires the corresponding response tensor $\varepsilon_{\alpha}$ to be symmetric,
and thus $\varepsilon_{\alpha}^{ca} = \varepsilon_{\alpha}^{ac}$ holds.
Therefore, one can understand that
\begin{align}
X^{tE}_{aa} = -X^{tE}_{cc} = -\sum_\alpha \varepsilon_{\alpha}^{ca}m^b_\alpha.
\end{align}
We confirm that this relationship also holds in the calculations for other olivine compounds presented in Appendix~\ref{app:A}.

Figure~\ref{fig:fig04}(d) displays the electric-field-direction dependence of the induced magnetic toroidal moment 
calculated using these coefficients. 
This plot clearly illustrates the highly anisotropic nature of this response. 
The prominent response along the $b$ component results in the large nonlinear magnetoelectric response discussed in the next subsection.

\subsection{Second-order magnetoelectric response}\label{sec:resultD}

The induced magnetic toroidal moment $\mathbf{t}$ activates a nonlinear magnetoelectric effect~\cite{Spaldin2008}.
Specifically, it induces antisymmetric components of the magnetoelectric tensor, 
resulting in a nonzero uniform magnetization
\begin{align}
\mathbf{m} = \frac{1}{4}
(\mathbf{m}_{\mathrm{A}} + \mathbf{m}_{\mathrm{B}} + \mathbf{m}_{\mathrm{C}} + \mathbf{m}_{\mathrm{D}})
\end{align}
in the direction perpendicular to both $\mathbf{E}$ and $\mathbf{t}$.
This represents a second-order magnetoelectric effect through a two-step coupling mediated by the magnetic toroidal moment.
In this phenomenon, crystal anisotropy plays an important role. 
In isotropic systems, the effective linear coupling between $\mathbf{E}$ and $\mathbf{t}$ 
can be written as $T_0\mathbf{E}\cdot\mathbf{t}$, resulting in $\mathbf{t} \parallel \mathbf{E}$, and therefore,
no magnetization is induced via this mechanism.
However, as illustrated in Fig.~\ref{fig:fig04}(d), 
the induced $\mathbf{t}$ in our anisotropic system is generally neither parallel nor antiparallel to $\mathbf{E}$. 
This obliqueness enables the emergence of an electric-field-induced magnetization perpendicular to $\mathbf{E}$ in 
this second-order magnetoelectric effect.

\begin{figure}[!tbp]
\centering
\includegraphics[trim = 0 0 0 0, width=\columnwidth,clip]{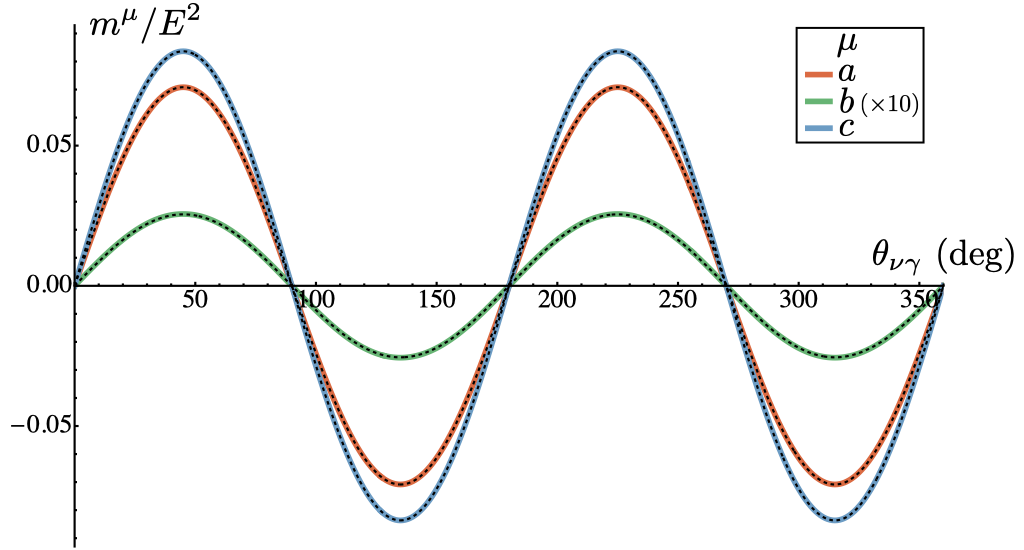}

\caption{\label{fig:fig05}
Electric-field angle dependence of the second-order magnetoelectric effect. 
A uniform magnetization $m^{\mu}$ along the $\mu$ axis is induced in the perpendicular direction to an applied electric field $\mathbf{E}$.
Solid lines are the numerical results obtained within the mean-field approximation, 
and dashed lines are fits to $X^{mEE}E^{2}\sin\theta\cos\theta$, 
where $X^{mEE}$ is the coefficient of the second-order magnetoelectric response and 
$\theta$ is the rotation angle of $\mathbf{E}$ in the measurement plane.
The horizontal axis is $\theta_{\nu\gamma}$ with $\{\mu\nu\gamma\}=\{abc\}$, $\{bac\}$, or $\{cab\}$,
corresponding respectively to the field rotations in the $ab$, $bc$, and $ac$ planes.
For visibility, the data for $\mu=b$ are multiplied by $10$.
}
\end{figure}

Figure~\ref{fig:fig05} shows the angular dependence of the field-induced magnetization $\mathbf{m}$ 
arising from this second-order magnetoelectric effect. 
As in the previous subsection, 
a sufficiently weak field with magnitude $E = 0.005$ is applied and rotated within the $ab$, $bc$, and $ac$ planes as Eqs.~\eqref{eq:Eab}--\eqref{eq:Eac}. 
The obtained angular dependences are analyzed using a fitting function proportional to $E^2 \sin\theta \cos\theta$,
which successfully reproduces the data for all field-rotation planes. 
This confirms that the induced perpendicular magnetization scales quadratically with the electric field. 
Defining the nonlinear magnetoelectric tensor by
\begin{align}
m^\mu = X^{mEE}_{\mu\nu\gamma} E^\nu E^\gamma,
\end{align}
the fits yield
\begin{align}
X^{mEE}_{abc} = X^{mEE}_{acb} = 0.1417, \\
X^{mEE}_{bca} = X^{mEE}_{bac} = 0.0051, \\
X^{mEE}_{cab} = X^{mEE}_{cba} = 0.1673;
\end{align}
all other components are zero. 
These results indicate that an antisymmetric toroidal-type magnetoelectric effect is induced by the electric field in this system, 
which would be verified in future experiments.

Finally, we remark on the nonlinear magnetoelectric response from a symmetry perspective. 
As mentioned in Sec.~\ref{sec:introduction}, the magnetic point group of the antiferromagnetic state in Co$_2$SiO$_4$ is $mmm$, indicating that this state is classified as an altermagnet~\cite{Yuan2021,Smejkal2022a,Smejkal2022b,Cheong2025}. 
More specifically,
the magnetic order considered in this study breaks both $\mathcal{T}$ and $\mathcal{PT}$ symmetry, and belongs to the S-type altermagnets, 
which allow higher-order off-diagonal magnetoelectric responses~\cite{Cheong2025}. 
In particular, a transverse even-order $E$-induced magnetization is symmetry-allowed in this S-type.
This feature is fully consistent with the present results.

\subsection{Other antiferromagnetic olivines}\label{sec:resultE}

\begin{figure}[!b]
\centering
\includegraphics[trim = 0 0 0 0, width=\columnwidth]{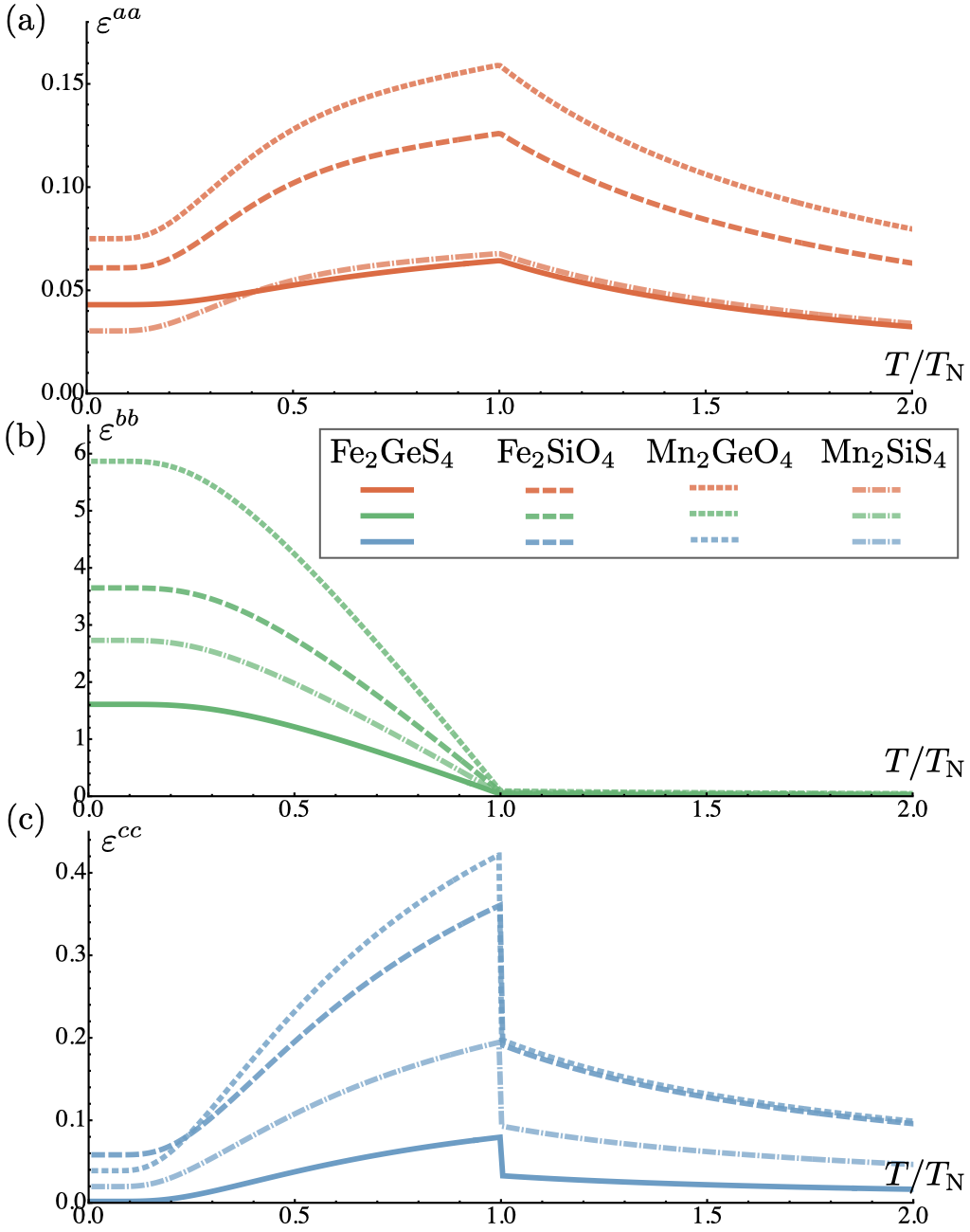}
\caption{\label{fig:fig06}
Temperature dependence of the dielectric constant
for the related olivine compounds
Fe$_2$GeS$_4$, Fe$_2$SiO$_4$, Mn$_2$GeO$_4$, and Mn$_2$SiS$_4$,
obtained within the mean-field approximation.
(a) $\varepsilon^{aa}$, (b) $\varepsilon^{bb}$, and (c) $\varepsilon^{cc}$ are shown.
}
\end{figure}

Building on the successful microscopic modeling of Co$_2$SiO$_4$, we extend our analysis to other olivine compounds. 
Several olivine compounds
have been reported to exhibit antiferromagnetic order qualitatively similar to that of Co$_2$SiO$_4$, 
including
Fe$_2$GeS$_4$~\cite{Junod1995},
Fe$_2$SiO$_4$~\cite{Lottermoser1986},
Mn$_2$GeO$_4$~\cite{White2012},
and
Mn$_2$SiS$_4$~\cite{Junod1995,Mandujano2023}. 
The magnetic ions in these compounds are Fe$^{2+}$ ($S=2$) and Mn$^{2+}$ ($S=5/2$),
differing from Co$^{2+}$ ($S=3/2$) in spin quantum number.
Although these materials share the same crystal symmetry,
the arrangements of ligands around the magnetic ions differ slightly.
Using the reported crystallographic data for each compound, 
Fe$_2$GeS$_4$~\cite{Vincent1976}, 
Fe$_2$SiO$_4$~\cite{Lottermoser1986},
Mn$_2$GeO$_4$~\cite{White2012},
and
Mn$_2$SiS$_4$~\cite{Lamarche1994},
we perform theoretical calculations following exactly the same procedure as for Co$_2$SiO$_4$.

Figure~\ref{fig:fig06} shows the temperature dependences of the dielectric constants
$\varepsilon^{aa}$, $\varepsilon^{bb}$, and $\varepsilon^{cc}$ for these olivine compounds,
corresponding to Figs.~\ref{fig:fig02}(a)--\ref{fig:fig02}(c) for Co$_2$SiO$_4$.
Although the magnitudes of the responses vary among compounds,
all of them exhibit qualitatively similar trends to those of Co$_2$SiO$_4$.
Within the present framework, Mn$_2$GeO$_4$ shows the largest overall response.
It should be noted, however, that 
the definition of the electric polarization operator used in this study sets all proportional coefficients $\lambda_{\alpha\mu}$ to unity for simplicity (see Sec.~\ref{sec:theory}).
For a quantitative comparison with experiments,
these coefficients should be determined based on experimental data.
Performing calculations corresponding to the analyses presented in Secs.~\ref{sec:resultC} and~\ref{sec:resultD}, 
we also confirm that all these related compounds exhibit qualitatively similar magnetoelectric behaviors to those in Co$_2$SiO$_4$ (see Appendix~\ref{app:A}).

Experimentally, 
unlike Co$_2$SiO$_4$, which shows a 
single magnetic transition, 
the other materials examined here undergo more complex phase transitions.
Specifically, Fe$_2$GeS$_4$ exhibits two successive magnetic transitions, with the intermediate-temperature phase sharing the same magnetic point group symmetry as the ordered phase of Co$_2$SiO$_4$~\cite{Junod1995}. 
Similarly, Fe$_2$SiO$_4$ also undergoes two transitions;
both ordered phases possess the same symmetry as that of Co$_2$SiO$_4$, 
while the low-temperature phase additionally exhibits a weak ferromagnetic moment~\cite{Tripoliti2023}. 
Mn$_2$GeO$_4$ also shows a similar antiferromagnetic order in an intermediate-temperature phase, despite exhibiting more complex behaviors through three successive magnetic transitions~\cite{White2012,Honda2012,Honda2014,Honda2017}. 
Finally, Mn$_2$SiS$_4$ undergoes an antiferromagnetic transition at low temperature, 
and a weak ferromagnetic order has been reported within a narrow temperature range just above the transition~\cite{Lamarche1994,Junod1995}, 
although a recent powder neutron diffraction study did not detect this additional order~\cite{Mandujano2023}.
These sequential phase transitions and complex magnetic behaviors are beyond the scope of
the present minimal model, indicating the need for further theoretical extensions.

\section{Summary and Perspectives}\label{sec:summary}

In this study, we have theoretically elucidated 
the microscopic origin of the magnetoelectric effect 
arising from an electric-field-induced magnetic toroidal moment in the antiferromagnetic olivine Co$_2$SiO$_4$.
Based on its crystal and magnetic structures, 
we constructed a minimal spin model that incorporates couplings of $S=3/2$ spins at the Co2 sites to an external electric field through the $d$-$p$ hybridization mechanism.
By applying the mean-field approximation to this model, 
we systematically evaluated the electric-field responses of the electric polarization, magnetic toroidal moment, and magnetization.
First, our theoretical calculations revealed anisotropic dielectric anomalies along the principal crystallographic axes at the antiferromagnetic transition temperature.
These trends qualitatively reproduce the experimental observations, 
demonstrating that the present minimal model provides a valid microscopic framework for understanding the magnetoelectric effect in Co$_2$SiO$_4$.
In addition, we showed that the antiferromagnetic order is accompanied by the emergence of magnetic toroidal monopoles,
which enable a linear response of the magnetic toroidal moment under an applied electric field.
The response is found to be anisotropic, with
both the magnitude and orientation of the induced moment strongly depending on the field direction.
Furthermore, 
our calculations clarified that a uniform magnetization is electric-field-induced via the magnetic toroidal moment as a second-order magnetoelectric effect,
which is symmetry-allowed within its altermagnet classification.
Finally, analogous analyses for related antiferromagnetic olivines demonstrated qualitatively similar behaviors to those of Co$_2$SiO$_4$, 
suggesting that the electric-field-induced magnetic toroidal phenomena are universally realized in a broad family of olivine-type antiferromagnets.

Future work should first extend the present model by including the Co1 sites,
which have been neglected here for simplicity.
Such an extension is expected to provide a more complete understanding of 
the detailed magnetic structure and magnetoelectric responses.
In addition, several of the related compounds exhibit multiple or complex magnetic transitions, 
with some magnetic phases sharing the same symmetry with that of Co$_2$SiO$_4$ and classified as altermagnetic states.
Clarifying the dielectric and toroidal responses in these phases and phase transitions between them
will be essential for uncovering the intrinsic nature of magnetoelectric coupling in olivine compounds.

From the viewpoint of antiferromagnetic domain control, 
the existence of magnetic toroidal monopoles $T_0$ is particularly noteworthy.
In systems with a finite $T_0$, not only the coupling term $E^\mu T^\mu$ but also symmetry-allowed terms 
such as  $\mathbf{m} \cdot (\nabla \times \mathbf{E}) \sim T_0\,\mathbf{m}\cdot(\partial_t \mathbf{B})$ can appear~\cite{Hayami2023}.
This implies the possibility of aligning and controlling antiferromagnetic domains through time-dependent magnetic fields.
Such novel magnetoelectric responses mediated by toroidal monopoles 
may open new routes for information-writing mechanisms in antiferromagnetic devices
and thus hold promise for future technological applications.

\begin{acknowledgments}
This work was supported by Japan Society for the Promotion of Science (JSPS) KAKENHI Grant 
Nos.~JP22K03509, JP25H00392, and JP25H01247.
\end{acknowledgments}

\renewcommand{\thesection}{\Alph{section}}
\renewcommand{\thesubsection}{\Alph{section}.\arabic{subsection}}
\renewcommand{\thesubsubsection}{\Alph{section}.\arabic{subsection}.\arabic{subsubsection}}
\appendix


\section{
Theoretical results for other antiferromagnetic olivines
}\label{app:A}

\begin{figure}[t]
\centering
\includegraphics[trim = 0 0 0 0, width=\columnwidth]{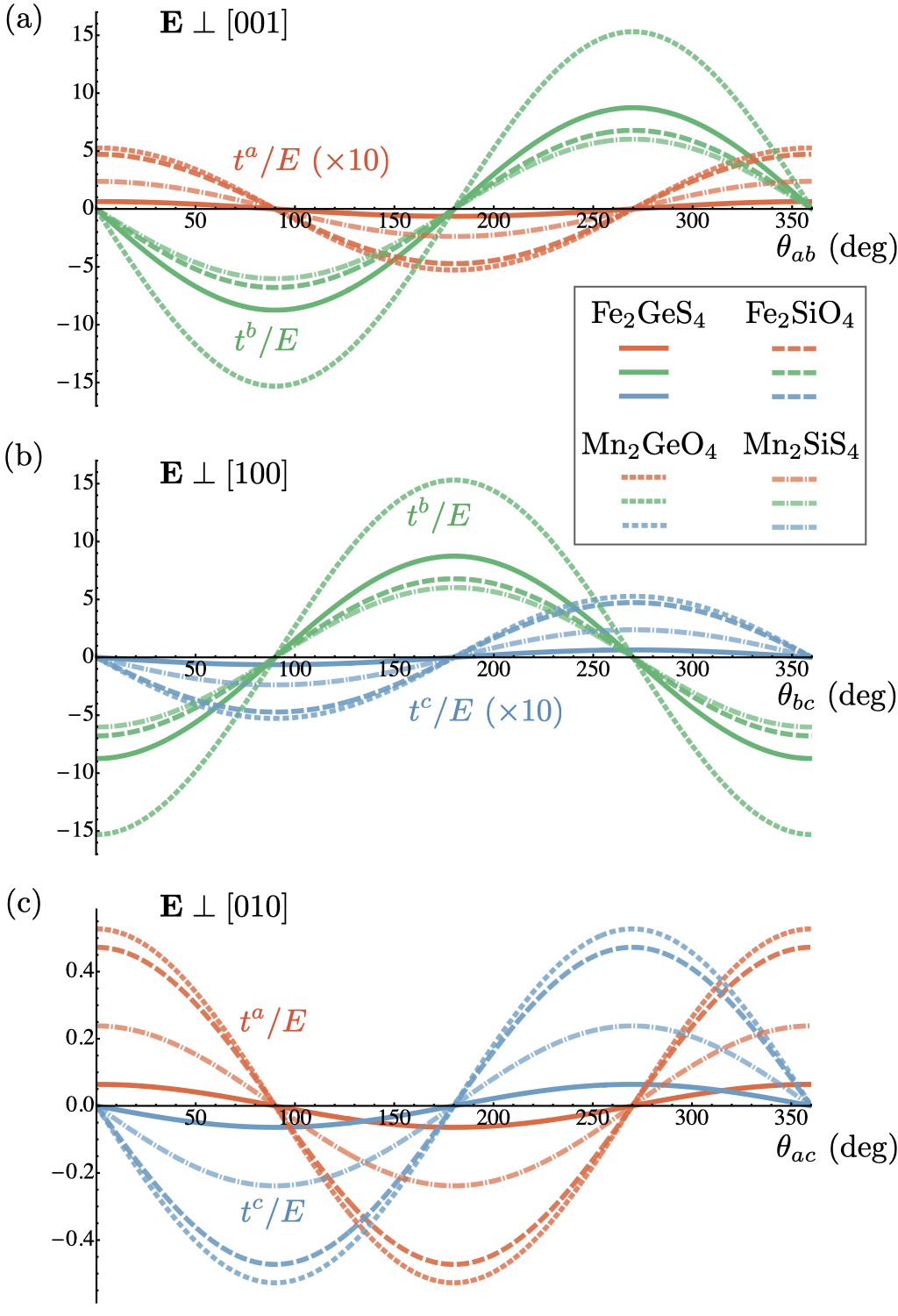}
\caption{\label{fig:fig07}
Electric-field-induced magnetic toroidal moments
for the related olivine compounds, 
Fe$_2$GeS$_4$, Fe$_2$SiO$_4$, Mn$_2$GeO$_4$, and Mn$_2$SiS$_4$, 
obtained within the mean-field approximation:
(a) $t^a$ and $t^b$ for $\mathbf{E}\perp c$,
(b) $t^b$ and $t^c$ for $\mathbf{E}\perp a$,
and (c) $t^a$ and $t^c$ for $\mathbf{E}\perp b$.
}
\end{figure}

\begin{figure}[t]
\centering
\includegraphics[trim = 0 0 0 0, width=\columnwidth]{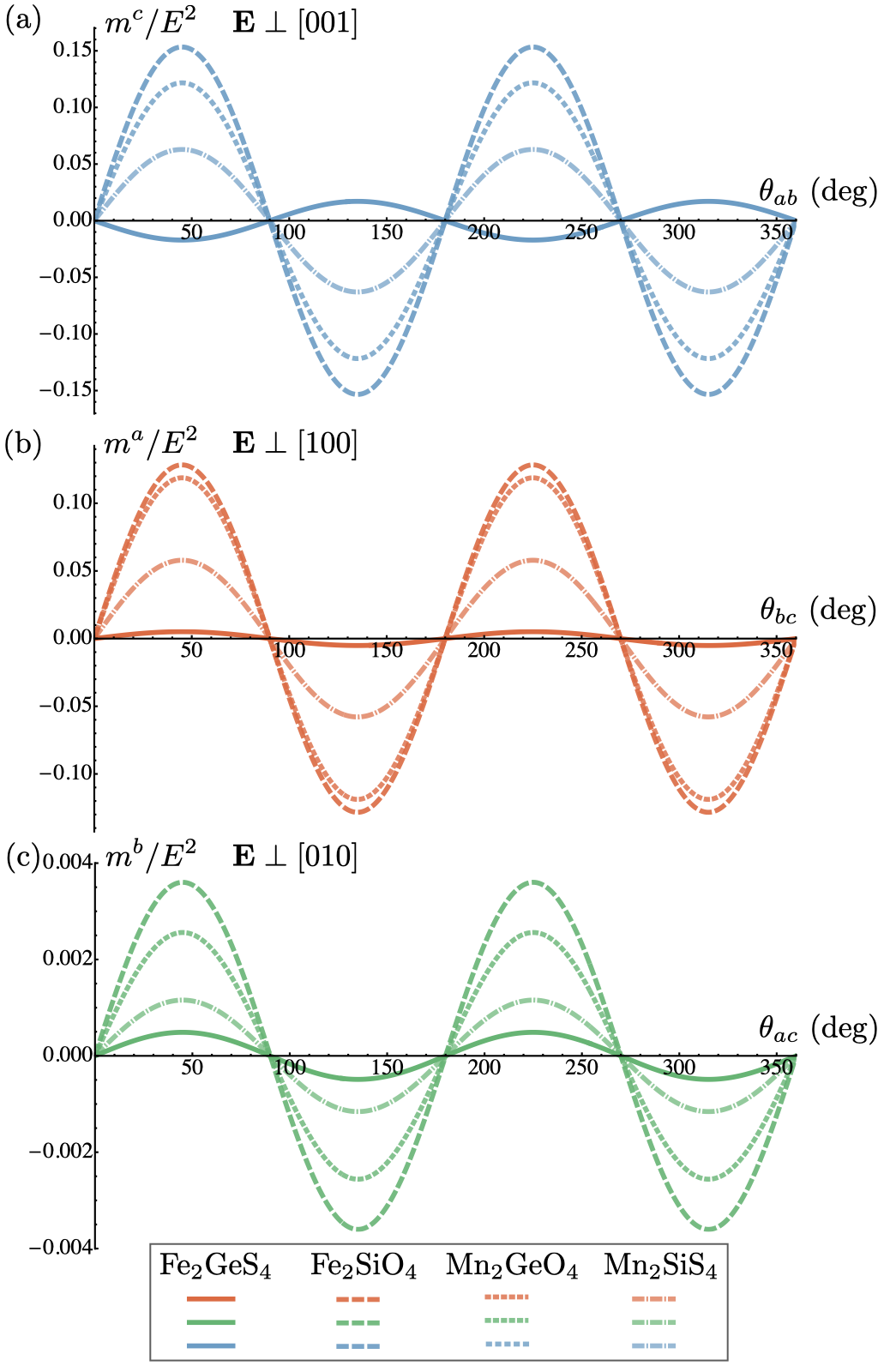}
\caption{\label{fig:fig08}
Electric-field angle dependence of the second-order magnetoelectric-induced uniform magnetization $m^{\mu}$.
Results are shown for the related olivine compounds, 
Fe$_2$GeS$_4$, Fe$_2$SiO$_4$, Mn$_2$GeO$_4$, and Mn$_2$SiS$_4$,
obtained within the mean-field approximation:
(a) $m^c$ for $\mathbf{E}\perp c$,
(b) $m^a$ for $\mathbf{E}\perp a$,
and (c) $m^b$ for $\mathbf{E}\perp b$.
}
\end{figure}

Figures~\ref{fig:fig07} and \ref{fig:fig08} show the electric-field-induced magnetic toroidal moments and uniform magnetization, respectively,
for the related olivine compounds, obtained through the mean-field calculations for the model in Eq.~\eqref{eq:H}. 
These results, corresponding to those in Figs.~\ref{fig:fig04} and \ref{fig:fig05}  for Co$_2$SiO$_4$, indicate qualitatively similar $T_0$-related magnetoelectric responses across all compounds.
We note that, in Fig.~\ref{fig:fig08}, 
$m^c$ induced by $\mathbf{E}$ in the $ab$ plane for Fe$_2$GeS$_4$ exhibits an opposite sign compared with those of the other compounds. 
We do not attribute physical significance to this sign reversal,
as it likely depends on the choice of the coefficient $\lambda_{\alpha\mu}$, 
which is set to unity in the present study. 


\clearpage

\bibliography{draft} 

\end{document}